\documentclass[12pt,a4paper]{article}
\usepackage[latin1]{inputenc}
\usepackage{amsmath}
\usepackage{placeins}
\usepackage{url}
\usepackage{natbib}
\usepackage{color,xcolor,setspace,geometry}
\setcitestyle{authoryear,open={(},close={)}}
\usepackage{cite}
\usepackage{soul}
\usepackage{bbm}
\usepackage{dsfont}
\soulregister\citep7
\soulregister\cite7
\soulregister\citet7
\soulregister\citealp7
\soulregister\ref7
\usepackage{amsfonts}
\usepackage{arydshln}
\usepackage{tikz}
\usetikzlibrary{shapes}
\usetikzlibrary{plotmarks}
\usetikzlibrary{tikzmark,matrix,calc}
\usetikzlibrary{arrows.meta,calc,decorations.markings,math,arrows.meta}
\usepackage{amssymb}
\usepackage{multirow}
\usepackage{todonotes}
\usepackage{graphicx}
\begin{document}
\begin{spacing}{1.35}	

\title{The reaction to news in live betting}

\author{Marius \"Otting\thanks{Bielefeld University} , Rouven Michels$^*$,
Roland Langrock$^*$, Christian Deutscher$^*$}
\date{}

\maketitle

\begin{abstract} 
Sports betting markets have grown very rapidly recently, with the total European gambling market worth 98.6 billion euro in 2019. Considering a high-resolution (1 Hz) data set provided by a large European bookmaker, we investigate the effect of news on the dynamics of live betting. 
In particular, we consider stakes placed in a live betting market during football matches. 
Accounting for the general market activity level within a state-space modelling framework, we focus on the market's response to events such as goals (i.e.\ major news), but also to the general situation within a match such as the uncertainty about the outcome. Our results indicate that markets might overreact to recent news, confirming cognitive biases known from psychology and behavioural economics.
\end{abstract}

\section{Introduction}

In the past decade, sports betting markets have grown very rapidly. For 2019, the European Gaming and Betting Association reported a gross gaming revenue --- that is, the amount of money bookmakers generate --- of about 98 billion euro. For comparison, as of 2010, the gross gaming revenue was reported as 6 billion euro only. The total amount of stakes placed has increased in recent years, and this concerns especially the market for bets placed during a match --- the so-called live betting market (sometimes also referred to as in-game betting or in-play betting). Whereas in the past the vast majority of bets were placed before the start of a match, nowadays the share of live betting accounts for about 55\% of the overall betting volume \citep{europeanbetting}.

In general, betting markets are similar to financial markets, as a bet on a team can be compared to buying a stock of a company \citep{sauer1998economics}. In contrast to betting markets, financial markets early on drew interest from empirical research, beginning with the study of the dynamics of stock prices \citep{bachelier}. Other studies have predicted stock returns in efficient markets \citep{balvers1990predicting}, analysed the drivers of stock returns \citep{hou2011factors}, investigated how different news (different in magnitude, type) affect stocks and searched for leverage effects \citep{chua2019information}, investigated the impact of panic-laden news (derived from news to the coronavirus outbreak) on the volatility of the financial market \citep{haroon2020covid}, and analysed dependencies between bank's operational losses and event types, including macroeconomic, financial, and firm-specific factors \citep{hambuckers2018understanding}, to name but a few. 
Financial time series data are typically driven by not necessarily directly observable states such as a market's activity level, the nervousness of the financial market, the state of the economy, or the regulatory environment. In many of the corresponding analyses, it thus makes conceptual sense to consider state-space (SSMs) or related models for relating the observed quantity of interest to underlying (latent) states (see, e.g., \citealp{svmodels}, \citealp{ALANASWAH20111073}, \citealp{warne2017marginalized}, and \citealp{barra2017joint}).

Compared to financial markets, live betting markets are in fact better suited for analysing the effect of news on a market's activity level, (i) since bets have a well-defined endpoint after which their value becomes observable \citep{thaler1988anomalies}, 
(ii) as major news such as goals refer to a single identifiable point in time, which is very rarely the case for news on financial markets, 
and (iii) as news in a live betting market become immediately available to all market participants due to the live coverage on the internet \citep{Jarrell}. 
Despite this excellent setting for investigations into how market participants and markets respond to news, the respective dynamics have hardly been investigated to date. 
\citet{gil}, \citet{choi2014role}, and \citet{croxson2014information} investigate the efficient market hypothesis using data from betting exchanges, where bettors bet against each other rather than bookmakers. Specifically, these studies analyse prices (i.e.\ betting odds) directly after goals were scored to check whether betting markets immediately incorporate such events into their prices.

In this contribution, we consider a high-resolution (1 Hz) data set covering stakes placed (i.e.\ the amount of money) in a live betting market, which was provided by a large European bookmaker. 
We use this unique data set to investigate betting dynamics in a football live betting market. Specifically, these high-resolution data enable a fine-grained analysis of betting patterns, which differ with respect to the course of the match or teams' strength. 
The main focus of our analysis is on the market's response to news. In particular, we investigate how news affect the stakes placed in a live betting market. As previous studies on live betting markets focus on the analysis of betting odds and not on stakes, our analysis to our knowledge is the first to investigate stakes placed in a live betting market, considering data from a bookmaker instead of betting exchanges as it was done in previous studies.

News in football betting can be represented by several events. Our main focus is on the most important news in a football match, namely goals scored. Given the relatively low number of goals of about three per game, each single goal has a considerable impact on the match outcome. However, a goal scored by an underdog may have a different effect on betting behaviour than a fourth goal in a blowout win by a clear favourite. Hence, not all goals are equally important to bettors. In addition,
goals alone are not sufficient to reflect the complexity of match dynamics. 
When modelling the impact of news on stakes in a betting market, we thus also include news in a wider sense, considering any change in the uncertainty about the outcome.
To model the stakes placed, we consider SSMs which have proven to be a useful statistical tool in similar investigations of financial markets. In our case, the latent states serve for the general market activity level, such that we are able to model the market's response to news.

\section{Data}\label{chap:data}
Data for this contribution were provided by one of Europe's largest bookmakers and detail the betting activity during German Bundesliga matches with a resolution of 1 Hz. We focus on match outcome bets (i.e.\ home or away win) as they attract most interest from bettors. Because the data are sampled at irregular time intervals, we aggregate betting activity into 15-second intervals to obtain regular time series. Since we are not allowed to provide any information on the actual stakes, all stakes in this paper were multiplicated with a constant beforehand.

\begin{figure}[htb]
	\centering
	\includegraphics[scale=0.5]{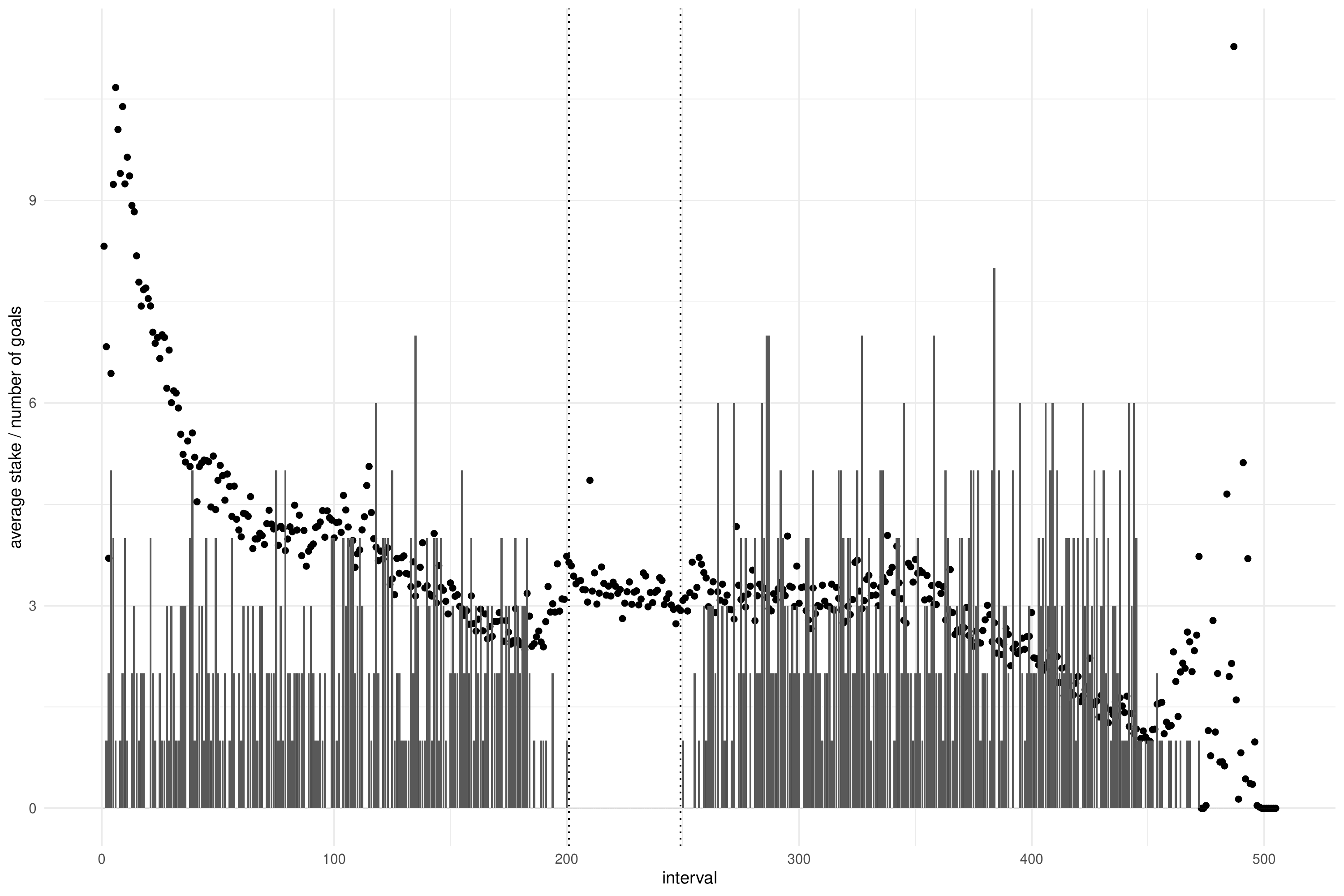}
	\caption{Stakes placed and number of goals per interval. The dots represent the average stakes placed in a given interval, and the bars show the number of goals per interval observed in our data. The dashed lines indicate halftime.}
	\label{img:stakes_and_goals}
\end{figure}

Our data comprises information on all German Bundesliga matches that took place during the 2018/19 season. Each of the 306 matches yields two time series, one for each team's bets, resulting in $612$ time series in total.
The time series of stakes are denoted by $\{ y_{n,t} \}$, with $\lceil n/2 \rceil$ indicating the match associated with the $n$--th series (home team if $n$ is odd), $n=1,\ldots,612$, and $t$ denoting the intervals, $t = 1, \ldots, T_n$ (each of length 15 seconds). Hence, each match is divided into at least 420 intervals, as a match consists of 90 minutes playing time, 15 minutes halftime, and a few (usually 1-5) minutes of injury time --- the number of intervals thus slightly varies across matches. In total, our data include 274,778 intervals (i.e.\ observations). Regarding the major news in a football match, i.e.\ the goals scored, we observe 969 goals in total, corresponding to an average of 3.2 goals per match. Figure \ref{img:stakes_and_goals} shows the average stakes placed in each interval (indicated by the dots), together with the number of goals observed in the interval (indicated by the bars). We observe relatively high stakes at the beginning of matches. For goals scored, we see that they are fairly even distributed within matches, with slightly more goals occurring in the second half.

\begin{figure}[htb]
	\centering
	\includegraphics[scale=0.5]{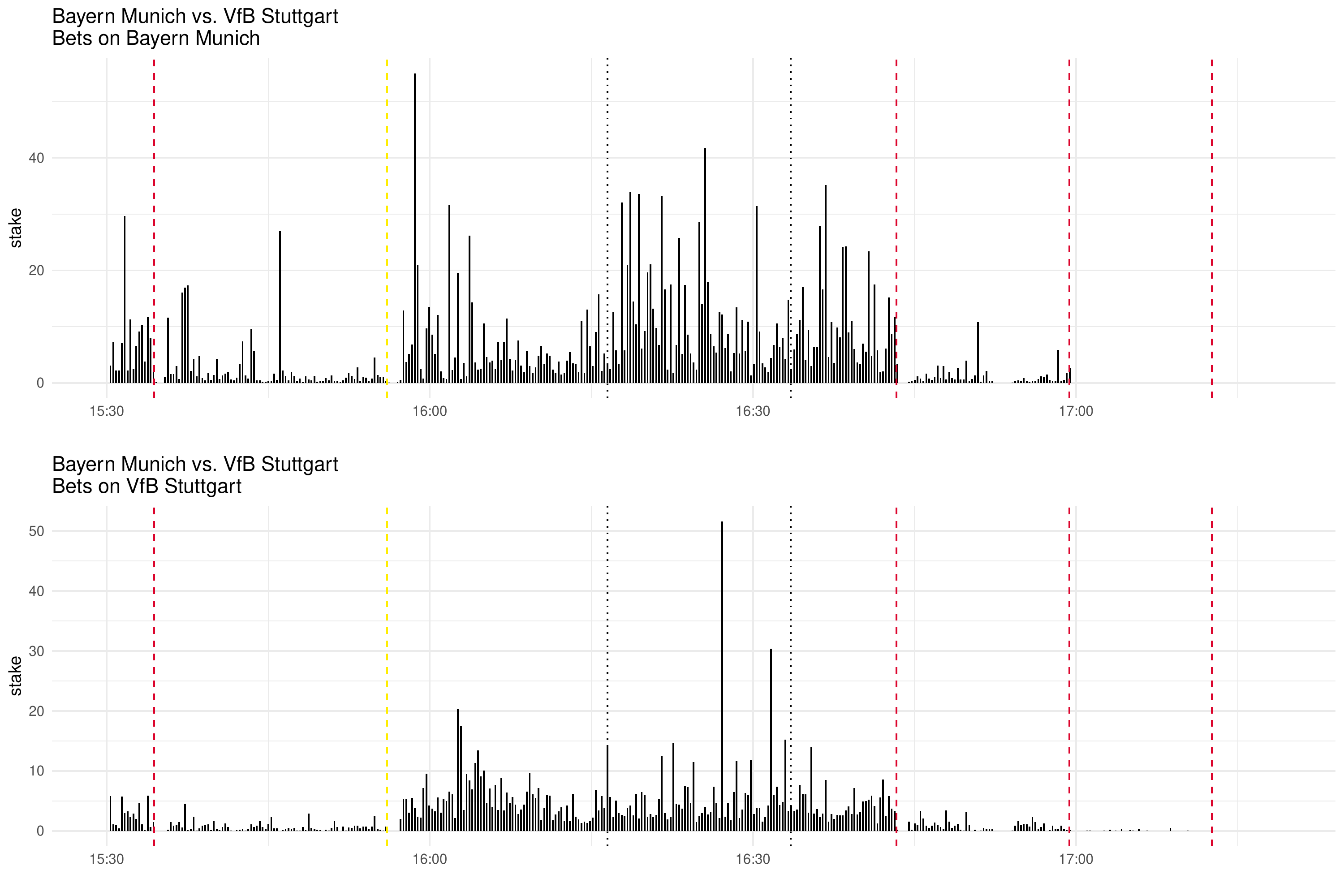}
	\caption{Time series of the stakes placed on a win of Bayern Munich for one example match from the data set (Bayern Munich vs.\ VfB Stuttgart). The vertical dashed lines denote goals scored by Bayern Munich (red lines) and VfB Stuttgart (yellow lines). Top: bets on Bayern Munich. Bottom: bets on VfB Stuttgart.}
	\label{img:grafik-dummy}
\end{figure}

To illustrate potential betting dynamics, Figure \ref{img:grafik-dummy} shows one example time series from the data set, corresponding to the bets placed on Bayern Munich in their match against VfB Stuttgart played on January 27, 2019.
It illustrates that dynamics in a live betting market likely depend on news such as goals and on the current situation in a match, i.e.\ the uncertainty about the outcome. To account for such characteristics, we include different covariates which represent characteristics of the underlying teams (such as strength of the teams) and events related to news during a football match. Our covariates, which are explained in further detail below, can be split into three categories, namely \textit{static}, \textit{market dynamics}, and \textit{news}.

We start with \textit{static} covariates associated with general match characteristics which remain constant throughout matches, namely
\begin{itemize}
    \item both teams' Elo ratings,
    \item and dummy variables for the different kick-off times. 
\end{itemize}
The covariates on the Elo ratings --- obtained from \url{http://clubelo.com/} --- serve as a suitable proxy for the teams' strength (see, e.g., \citealp{HVATTUM2010460}). The Elo ratings vary across matches, i.e.\ they account for the current form of a team, but remain constant within a single match. It seems intuitively plausible that teams with higher Elo ratings attract more bets, simply as better teams tend to attract more fans, many of whom prefer to bet on their favourite team (see, e.g., \citealp{feddersen2017sentiment} and \citealp{FeddersenNBA}). This is similar to sentiment traders in financial markets, whose investment decisions are beyond rational pricing (see e.g., \citealp{baker2007investor}, \citealp{brown2005investor} \citealp{braun2013national}). 
The kick-off time of the match, especially the day of the week on which the match is played, is included since previous studies investigating pre-game betting suggest that stakes vary across days of the week (see \citealp{HUMPHREYS2013376} and \citealp{deutscher2019demand}).
In our analysis, we thus consider dummy variables for the different kick-off times.

In addition to these static covariates, we include \textit{market dynamic} covariates related to market features, i.e.\ information on the market and general match status, via
\begin{itemize}
    \item a dummy variable indicating whether the market is open, 
    \item a dummy variable indicating whether the match is in its halftime break, 
    \item a covariate indicating the number of remaining intervals at interval $t$,
    \item and a covariate indicating the imbalance of the match at interval $t$ (based on the winning probabilities of both teams). 
\end{itemize}
As bettors are not able to bet on events if markets are closed, the first covariate simply indicates whether the market was open in interval $t$. We observe a closed market directly after certain events such as goals and red cards, since bookmakers need some time to adjust their odds. In Figure \ref{img:grafik-dummy}, we thus observe no bets placed immediately after goals were scored.
We further include a dummy variable indicating the halftime break as increased stakes during halftime have been reported in the literature \citep{croxson2014information}, and a covariate indicating the number of remaining intervals at interval $t$.
Moreover, Figure \ref{img:grafik-dummy} reveals that, for some periods, no bets are placed although the market is open. This phenomenon usually occurs when the match seems clearly decided, e.g.\ after Bayern Munich scored their third goal in the example time series. 
To account for this, we introduce the Gini coefficient at interval $t$. Based on the betting odds at interval $t$, we can simply retrieve the implied probabilities for a home win, a draw, and an away win (accounting for the bookmaker's margin). The resulting Gini coefficient based on such implied probabilities ranges between zero and one, where a Gini of one refers to a match which is decided, whereas a Gini of zero refers to a match which is completely balanced --- according to the bookmaker's odds.
In Figure \ref{img:grafik-dummy}, after Bayern Munich's second goal, the odds for a win of Stuttgart, for a draw, and for a win of Bayern Munich were 50, 13, and 1.03, respectively (equivalent to probabilities of 0.0187, 0.072, 0.909) and thus the corresponding Gini coefficient is obtained as 0.89. Since the match was almost decided at that point, we observe less bets. 

Finally, as we are primarily interested in analysing responses of bettors to news, we include the following \textit{news} covariates:
\begin{itemize}
    \item covariates measuring the time since the last goal of the team being analysed
    \item and a covariate measuring changes in the expected match outcome compared to the kick-off.
\end{itemize}
To account for changes in betting dynamics after goals were scored, we include covariates measuring the time since the last goal reflecting that bettors may be attracted to place a bet after a goal has been scored. Specifically, we categorise our covariate depending on the odds right before the goal. We distinguish between \textit{surprising} goals, with odds above $4.00$ at the time of the goal, corresponding to a probability to win the match (without bookmaker's margin) of below 25\%, \textit{slightly surprising} goals (odds between $2.00$ and $4.00$, win probability between 25\% and 50\%), and \textit{unsurprising} goals (odds below $2.00$, win probability above 50\%). Unsurprising (surprising) goals refer to teams whose winning probability is rather high (low) at that time of the goal. We thus consider the odds as a valid proxy for the surprise-level of a scored goal.

Throughout the match, the odds and hence the implied winning probabilities change due to actions on the pitch. To account for uncertain courses of a match which may cause increased stakes, we introduce a further covariate, called \textit{ginidiff} in the following, capturing such changes by subtracting the current Gini at interval $t$ from the Gini at the beginning. A decreasing \textit{ginidiff} value coincides with a rather uncertain course of the match, for example after VfB Stuttgart equalised against Bayern Munich (see Figure \ref{img:grafik-dummy}). Here, the \textit{ginidiff} value dropped from 0.029 to -0.027, which means that after the goal the bookmaker considered the winning chances to be more balanced than at the beginning. Vice versa, an increasing \textit{ginidiff} coincides with a course of the match associated with decreasing outcome uncertainty, as indicated by a \textit{ginidiff} value of 0.1765 after Bayern Munich scored their fourth goal. Table \ref{tab:covariates} briefly summarises all covariates considered.

\begin{table}[!htbp] \centering 
  \caption{Descriptive statistics of the variable analysed (stake), as well as the \textit{static}, \textit{market dynamic}, and \textit{news} covariates.} 
  \label{tab:covariates} 
\scalebox{0.8}{
\begin{tabular}{@{\extracolsep{5pt}}lcccc} 
\\[-1.8ex]\hline 
\hline \\[-1.8ex] 
 & \multicolumn{1}{c}{mean} & \multicolumn{1}{c}{st.\ dev.} & \multicolumn{1}{c}{min.} & \multicolumn{1}{c}{max.} \\ 
\hline \\[-1.8ex] 
stake & 3.755 & 8.347 & 0 & 1027 \\ 
\hdashline
Elo ratings (\textit{eloteam} and \textit{eloopp}) & 1675 & 104.2 & 1469 & 1980 \\ 
matches played on Friday (\textit{friday}) & 0.095 & -- & 0 & 1 \\ 
matches played on Saturday afternoon (\textit{saturdayaft}) & 0.513 & -- & 0 & 1 \\ 
matches played on Saturday evening (\textit{saturdayeve}) & 0.101 & -- & 0 & 1 \\ 
matches played on Sunday (\textit{sunday}) & 0.216 & -- & 0 & 1 \\ 
matches played on weekday (reference category) & 0.075 & -- & 0 & 1 \\ 
\hdashline
status of the market (\textit{open}) & 0.908 & -- & 0 & 1 \\ 
halftime (\textit{halftime}) & 0.149 & -- & 0 & 1 \\ 
remaining intervals (\textit{int}) & 210.1 & 121.7 & 0 & 472 \\ 
\hdashline
intervals since last goal (\textit{goalteam}) & 47.85 & 79.22 & 0 & 422 \\ 
Gini coefficient (\textit{gini}) & 0.544 & 0.297 & 0 & 0.993 \\ 
difference between the Gini at interval $t$ and interval 1 (\textit{ginidiff}) & 0.205 & 0.292 & -0.794 & 0.916 \\
\hline \\[-1.8ex] 
\end{tabular}}
\end{table} 

\section{Modelling stakes in a live betting market}\label{chap:models}

Figure \ref{img:grafik-dummy} indicates clear serial correlation in the time series to be modelled, with extended periods of low (e.g.\ before 16:00) but also of high betting activity (e.g.\ around 16:30). This example time series further highlights that betting patterns vary according to the course of a match, and in particular in response to important events, namely goals: when a team has a clear lead and the match outcome is fairly certain (e.g.\ after Munich's third goal), less money is bet compared to a situation where the score is even (e.g.\ after Stuttgart's goal).
To capture the serial correlation, which to some extent is induced by news such as goals,
we consider covariate-driven SSMs, which formalise the idea of a market progressing through different phases, with an underlying state process measuring the (latent) level of market activity. 

If the state process within the SSM was discrete, say with only two states, then these could correspond to a highly active market (e.g.\ in a match where the lead changes several times) and a rather inactive market (e.g.\ in a match with an early clear lead of the favourite), respectively. However, while discrete states may be easy to interpret, the use of a continuous-valued state process provides more flexibility as a corresponding model can capture gradual changes in the underlying market activity level, which makes more sense conceptually as there is no clear justification for a finite number of market activity levels. This reasoning is analogous to that of corresponding modelling approaches in finance, where share returns are typically modelled via stochastic volatility models, with a continuous-valued state process (see, e.g., \citealp{ai2007maximum,langrock2012some,barra2017joint}). We thus consider a discrete-time SSM with continuous state space to model the time series of stakes placed. 

We first formulate a baseline SSM without covariates. Subsequently, we will step-by-step extend this baseline model to better capture all relevant patterns observed in the data, thereby gradually increasing the model's complexity. The final model formulation will contain all covariates introduced above.

\subsection{A baseline model}\label{chap:baselinemodel}

Our baseline model is a simple SSM with the observed
state-dependent process $\{ y_{n,t} \}$ --- $y_{n,t}$ indicating the stakes placed during interval $t$ of the $n$--th time series --- assumed to be driven by an unobserved state process $\{ g_{n,t} \}$. For notational simplicity, we will drop the index $n$ in the following. In general, SSMs are of the following form:
\begin{align}
\label{eq:allgemein}
    y_t = a(g_t, \epsilon_t), \ g_t = b(g_{t-1}, \eta_t), \ t = 1, . . ., T, 
\end{align}
where $a(.,.)$ and $b(.,.)$ are known functions, and $\epsilon_t$ and $\eta_t$ are the observation and the process errors, respectively.
As our response variable $y_t$ is continuous-valued and non-negative, with some zeros, we model it using the zero-adjusted gamma distribution (ZAGA), which is a mixture of the standard gamma distribution and a point mass on zero (see, e.g., \citealp{rigby2019distributions}, and \citealp{TONG2013548}, for an application):
\begin{center}
    $y_t \sim \text{ZAGA}(\mu_t, \sigma, \pi_t), \,\,\,$ with
        $f(y_t) = \begin{cases} \pi_t, \text{ if } y_t=0; \\ (1 - \pi_t) h(y_t),  \text{ if } y_t>0, \end{cases}$
\end{center}
with $h(y_t)$ the density of the gamma distribution parameterised in terms of mean $\mu_t>0$ and standard deviation $\sigma>0$ (see \citealp{rigby2019distributions}).
As discussed in Section 2, after events such as goals, red cards, and penalty kicks the live betting market is closed, such that no bets can be placed. To accommodate this within our model, we include the dynamic information on whether the market is open ($open_t$ = 1) into our predictor for $\pi_t$,
setting $\pi_t=\pi$ if $open_t = 1$ and  $\pi_t=1$ otherwise (if $open_t$ = 0).

In the ZAGA distribution, $\sigma$ is assumed to be constant, while the state variable $\{g_t\}$, modelled as a first-order autoregressive process, is assumed to affect $\mu_t$ to account for the time-varying activity level of the market:
\begin{center}
$\mu_t = \exp (\alpha_0 + g_t)$, \\
$g_t = \phi g_{t-1} + \eta_t$.
\end{center} 
For the error term we assume $\eta_t \stackrel{\text{iid}}{\sim} \mathcal{N}(0,\sigma_g^2)$ with $\sigma_g > 0$ controlling the amount of variation in the market volatility. For the process $\{g_t\}$ to be stationary, $|\phi| < 1$ is required.
For the initial distribution, i.e.\ at the beginning of each match, we assume $g_1 \sim \mathcal{N} \bigl( 0,\sigma_g/\sqrt{1 - \phi^2}\bigr)$, i.e.\ that $g_t$ is in its stationary distribution at $t=1$. 
For $\phi = 0$, the SSM collapses to a regression model without any serial correlation. 
Otherwise, i.e.\ if $\phi$ differs from zero, we observe serial correlation in the state process --- i.e.\ in the market activity level. 

\subsection{Including covariates in the state-dependent process}
To improve the realism of our model and to investigate the drivers of the betting activity, as a next step we include the static covariates introduced above in the model for the amount of stakes placed. Expecting higher stakes to be placed on top teams, we include the Elo rating as a proxy for team strength. 
Finally, we include dummy variables for the kick-off times. We then have as the predictor for $\mu_t$:
\begin{align*}
    \mu_t = \exp \bigl( \alpha_0 &+ \alpha_1 eloteam + \alpha_2 eloopp \\&+\omega_1 friday + \omega_2 \textit{saturdayaft}  + \omega_3 saturdayeve + \omega_4 sunday + g_t \bigr).
\end{align*}
The reference category for the categorical kick-off time variable are weekdays on which matches are played (i.e.\ Tuesday and Wednesday).

Furthermore, Figure \ref{img:grafik-dummy} already indicated that almost no stakes are placed if the outcome of a match is fairly certain: both panels in Figure \ref{img:grafik-dummy} indicate that if a team has a clear lead, such as Bayern Munich's 3-1 lead (cf.\ the third red orthogonal line) against VfB Stuttgart, bettors effectively stop wagering money on either team. We thus include dynamic information regarding the actual market and general match status, specifically \textit{gini} and \textit{int}, into the predictor for $\pi_t$. The effect of the former is modelled using a quadratic term as we found this additional flexibility to be required to fully capture this relationship. We further include an interaction effect between the two covariates. 
The probability of a zero stake in interval $t$ hence is modelled as
$$
\pi_t = 
\text{logit}^{-1} (\gamma_0 + \gamma_1 gini_t + \gamma_2 gini_t^2 + \gamma_3 int_t + \gamma_4 gini_t \cdot int_t + \gamma_5 gini_t^2 \cdot int_t)
$$
if $\text{open}_t=1$, and $\pi_t =1$ otherwise, i.e.\ if $\text{open}_t=0$. 

\subsection{Including covariates in the state process}

For the final model formulation, we extend the AR(1) state process to an ARX(1) process, i.e.\ we include the news covariates introduced above in the unobserved state process $g_t$. 
The values of these news covariates vary during the match, such that we can investigate their effect on the market activity level:
\begin{center}
$g_t = \phi g_{t-1} + \beta_1 \textit{halftime}_t + \beta_2 \textit{ginidiff}_t + \sum_{j = 3}^5 \beta_j \frac{1}{goalteam^{(j)}_{t}} + \sigma_g \eta_t$, 
\end{center} where $j = 3$ correspond to \textit{suprising}, $j = 4$ to \textit{slightly} and $j = 5$ to \textit{unsurprising} goals.
The dummy variable \textit{halftime} simply serves to capture a potentially elevated market activity level during halftime. The effect of the course of the match, compared to kick-off, is captured by \textit{ginidiff}. A deviation from the expected outcome can hence be interpreted as news, e.g.\ if the clear underdog takes the lead. Furthermore, the occurrences of goals in a match are likely to elevate the market activity level. However, Figure \ref{img:grafik-dummy} indicates that such an effect vanishes rather quickly, with the effect of a goal being highest directly after the goal was scored. 
To account for this pattern, we include the covariates \textit{goalteam} 
covering the time since the last goal was scored as $\frac{1}{goalteam^{(j)}_{t}}$, where $j$ distinguishes between the three different types of goals as categorised based on the odds.
Before presenting the results, we will introduce how to conduct maximum likelihood estimation for the SSMs presented in this section.

\subsection{Maximum likelihood estimation}\label{chap:estimation}

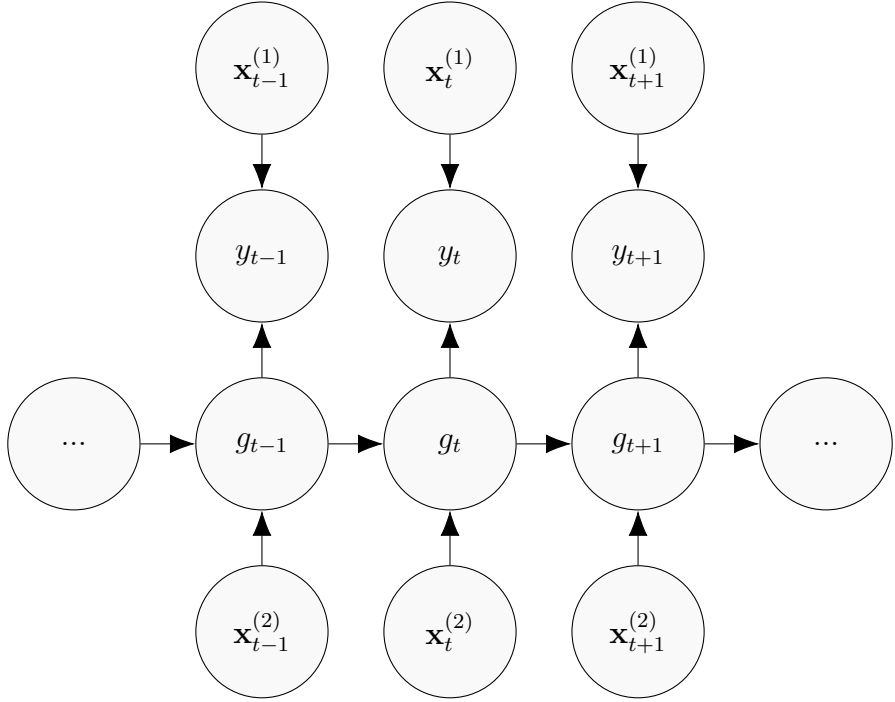
\begin{figure}[!htb]
    \centering
	\begin{tikzpicture}
	% nodes
	\node[circle,draw=black, fill=gray!5, inner sep=0pt, minimum size=50pt] (COV1) at (2, 0) {$\mathbf{x}^{(1)}_{t-1}$};
	\node[circle,draw=black, fill=gray!5, inner sep=0pt, minimum size=50pt] (COV2) at (4.5, 0) {$\mathbf{x}^{(1)}_{t}$};
	\node[circle,draw=black, fill=gray!5, inner sep=0pt, minimum size=50pt] (COV3) at (7, 0) {$\mathbf{x}^{(1)}_{t+1}$};
	\node[circle,draw=black, fill=gray!5, inner sep=0pt, minimum size=50pt] (COV4) at (2, -7.5) {$\mathbf{x}^{(2)}_{t-1}$};
	\node[circle,draw=black, fill=gray!5, inner sep=0pt, minimum size=50pt] (COV5) at (4.5, -7.5) {$\mathbf{x}^{(2)}_{t}$};
	\node[circle,draw=black, fill=gray!5, inner sep=0pt, minimum size=50pt] (COV6) at (7, -7.5) {$\mathbf{x}^{(2)}_{t+1}$};
	\node[circle,draw=black, fill=gray!5, inner sep=0pt, minimum size=50pt] (A) at (2, -5) {$g_{t-1}$};
	\node[circle,draw=black, fill=gray!5, inner sep=0pt, minimum size=50pt] (A1) at (-0.5, -5) {...};
	\node[circle,draw=black, fill=gray!5, inner sep=0pt, minimum size=50pt] (B) at (4.5, -5) {$g_{t}$};
	\node[circle,draw=black, fill=gray!5, inner sep=0pt, minimum size=50pt] (C) at (7, -5) {$g_{t+1}$};
	\node[circle,draw=black, fill=gray!5, inner sep=0pt, minimum size=50pt] (C1) at (9.5, -5) {...};
	\node[circle,draw=black, fill=gray!5, inner sep=0pt, minimum size=50pt] (Y1) at (2, -2.5) {$y_{t-1}$};
	\node[circle,draw=black, fill=gray!5, inner sep=0pt, minimum size=50pt] (Y2) at (4.5, -2.5) {$y_{t}$};
	\node[circle,draw=black, fill=gray!5, inner sep=0pt, minimum size=50pt] (Y3) at (7, -2.5) {$y_{t+1}$};
	\draw[-{Latex[scale=2]}] (A)--(B);
	\draw[-{Latex[scale=2]}] (B)--(C);
	\draw[-{Latex[scale=2]}] (A1)--(A);
	\draw[-{Latex[scale=2]}] (C)--(C1);
	\draw[-{Latex[scale=2]}] (A)--(Y1);
	\draw[-{Latex[scale=2]}] (B)--(Y2);
	\draw[-{Latex[scale=2]}] (C)--(Y3);
	\draw[-{Latex[scale=2]}] (COV1)--(Y1);
	\draw[-{Latex[scale=2]}] (COV2)--(Y2);
	\draw[-{Latex[scale=2]}] (COV3)--(Y3);
	\draw[-{Latex[scale=2]}] (COV4)--(A);
	\draw[-{Latex[scale=2]}] (COV5)--(B);
	\draw[-{Latex[scale=2]}] (COV6)--(C);
	\end{tikzpicture}
\caption{Dependence structure of the SSMs considered: every observation $y_{t}$ is generated by the conditional distribution implied by state process $g_{t}$, and the state process satisfies the Markov property. The vectors $\mathbf{x}^{(1)}_{t}$ and $\mathbf{x}^{(2)}_{t}$ include the covariates introduced in Section \ref{chap:data}, which affect the state-dependent and the state process, respectively.}
\label{fig:HMM}
\end{figure}

To evaluate the likelihood of the SSMs presented above, we will use a combination of numerical integration and recursive computing, as first suggested by \citet{Kitagawa}. Exploiting the model's dependence structure --- i.e.\ the Markov property of the state process and the conditional independence of the observations (see Figure \ref{fig:HMM}) --- we first rewrite the likelihood as follows:
\begin{align*}
\mathcal{L}(\boldsymbol\theta) &= f(y_1, \ldots, y_T) \\
&= \int \ldots \int f(y_1, \ldots, y_T, g_1, \ldots, g_T) \ \text{d}g_T \ldots \text{d}g_1 \\
&= \int \ldots \int f(y_1, \ldots, y_T | g_1, \ldots, g_T) f(g_1, \ldots, g_T)\ \text{d}g_T \ldots \text{d}g_1 \\
&= \int \ldots \int f(g_1)f(y_1| g_1) \prod\limits_{t = 2}^{T} f(g_t| g_{t-1}) f(y_t|g_t)\ \text{d}g_T \ldots \text{d}g_1,
\end{align*}
with $\boldsymbol\theta$ denoting the vector comprising all model-defining parameters. For some of the models defined above, the conditional densities in this likelihood depend on exogenous variables such as \textit{ginidiff} or \textit{halftime}, which for simplicity is not made explicit in the notation. 

Following \citet{zucchini}, we then use numerical integration via a simple midpoint rule, effectively discretising the state space such that market activity levels are aggregated into intervals $B_i = [b_{i-1},b_i]$, $i = 1, ..., m$, all of length $h = (b_m-b_0)/m$: 
$$ 
\mathcal{L}(\boldsymbol\theta) \approx h^T \sum\limits_{i_1 = 1}^m ... \sum\limits_{i_T = 1}^m f(b_{i_{1}}^{\star})f(y_1|g_1 = b_{i_{1}}^{\star}) \prod\limits_{t = 2}^{T} f(b_{i_{t}}^{\star}|g_{t-1} = b_{i_{t-1}}^{\star})f(y_t|g_t = b_{i_{t}}^{\star}),
$$
with $b_1^{\star},...,b_m^{\star}$ denoting the midpoints of the $m$ intervals. Even though this expression involves $m^T$ summands, where both $m$ and $T$ will be fairly large, it can be calculated at computational cost $\mathcal{O}({Tm^2})$ (only) via the recursive forward algorithm. For this, the likelihood is regarded as that of an $m$-state hidden Markov model (HMM), such that the corresponding machinery can be applied. Specifically, we consider an $m$-state HMM with
\begin{itemize}
\item initial distribution $\boldsymbol{\delta}$ such that $\delta_i = hf(b_i^{\star})$, $i = 1, \ldots, m$ 
(i.e.\ the probability of the HMM starting in state $i$ is the approximate probability of the market activity level at time~1 lying  in the interval $B_i$);
\item $m\times m$ transition probability matrix $\boldsymbol{\Gamma}^{(t)} = (\gamma_{ij}^{(t)})$ such that 
$\gamma_{ij}^{(t)} = hf(b_j^{\star}|g_{t-1} = b_{i}^{\star})$,  $i,j = 1, \ldots, m$ 
(i.e.\ the probability of the HMM switching from state $i$ to state $j$ at time $t$ is the approximate probability that a market activity level of $b_{i}^{\star}$ at time $t-1$ is followed by some activity level in $B_j$ at time $t$);
\item state-dependent density of the stakes $y_t$, given state $i$ at time $t$, as the ZAGA distribution's density $f(y_t|g_t=b_{i}^{\star})$, $i=1,\ldots,m$.
\end{itemize}
The likelihood of this HMM is \textit{precisely} the approximate likelihood of the SSM derived above via numerical integration. Crucially, this approximation can be made arbitrarily accurate by increasing $m$. Several previous studies have demonstrated that virtually exact maximum likelihood estimates are obtained when choosing $m\geq 50$, with $m=100$ typically already presenting a conservative choice (see, e.g., \citealp{langrock2013maximum}, \citealp{abanto2017maximum}, \citealp{mews2020maximum}), 
provided that the range over which the state process is integrated, $[b_0,b_m]$, covers nearly all mass of the distribution. The accuracy of the approximation can easily be checked by increasing $m$ as well as the width of the range (i.e.\ $b_m-b_0$) --- when this results only in negligible changes in the estimates and the maximum of the likelihood function, then the approximation is virtually exact. 

Summarising the state-dependent densities at time $t$ in the $m \times m$ diagonal matrix $\mathbf{P}(y_t)$ with the $i$--th entry $f(y_t|g_t=b_{i}^{\star})$, the likelihood of the HMM, and hence the approximate SSM likelihood, can now be calculated recursively as
\begin{center}
    $\mathcal{L}_{\text{approx}} = \boldsymbol{\delta} \mathbf{P}(y_1)\boldsymbol{\Gamma} \mathbf{P}(y_2)\boldsymbol{\Gamma} \mathbf{P}(y_3) \dots \boldsymbol{\Gamma} \mathbf{P}(y_{T-1})\boldsymbol{\Gamma} \mathbf{P}(y_T) \mathbf{1}$,
\end{center}
with column vector $\mathbf{1}=(1,\ldots,1)' \in \mathbb{R}^m$ \citep{zucchini}. This is the likelihood of a single time series. For the longitudinal data set considered here, we assume independence between the observed stakes on the outcomes of the different teams and matches, such that we obtain the likelihood for the full data set as the product of the $612$ individual likelihoods.
To obtain parameter estimates, the approximate SSM likelihood is maximised numerically in R using the function \texttt{nlm()}, subject to routine technical challenges such as parameter constraints (addressed by us using reparameterisation to unconstrained parameters), numerical underflow of the likelihood function (addressed using a scaling strategy to calculate the log-likelihood), and local maxima (addressed by trying out many different random initial values in the numerical maximisation).

\section{Results}\label{chap:results}

\subsection{Baseline model}
When fitting the baseline model, we use $m=150$ and $-b_0 = b_m = 4$ in the likelihood approximation. Table \ref{tab:resultsBaseline} displays the results. The estimate for $\phi$ is fairly close to 1 ($\hat{\phi} = 0.985$), indicating strong serial correlation in the underlying market activity level. 
Indeed, compared to a model without a state process, the AIC clearly favours the state-space formulation ($\Delta \text{AIC} = 217117$). 
With $\hat{\pi} = 0.074$, the estimated probability for observing a stake of zero in an open market matches the corresponding empirical proportion of $0.072$. 

\begin{table}[htb]
\centering
\caption{Parameter estimates with 95\% confidence intervals for the baseline model. 
}\vspace{0.3em}
\label{tab:resultsBaseline}
\begin{tabular}{rccc}
  \hline
 parameter & estimate & \multicolumn{1}{c}{95\% CI} \\ 
  \hline
$\phi$ & 0.985 & [0.984;\,0.986] \\ 
$\sigma_g$ & 0.227 & [0.223;\,0.230] \\
$\alpha_0$ & 0.642 & [0.590;\,0.693] \\
$\sigma$ & 0.872 & [0.870;\,0.873] \\ 
$\pi$ & 0.074 & [0.072;\,0.075] \\ 
   \hline
\end{tabular}
\end{table}

\subsection{Covariates in the state-dependent process}
Table \ref{tab:resultscovariates1} displays the results obtained for the model accommodating additional covariates in the stakes' (ZAGA) distribution, in particular related to team heterogeneity and uncertainty about the match outcome. For the estimated effects on the stakes' time-dependent mean $\hat{\mu}_t$, we see that higher quality teams tend to attract more money ($\hat{\alpha}_1 > 0$), whereas less money is placed on teams which face a strong opponent ($\hat{\alpha}_2 < 0$). 
Stakes also generally tend to be higher whenever there are no parallel matches --- indicated by a negative effect for the games played at Saturday afternoon ($\hat{\omega}_2 < 0$) --- which is in line with the existing literature (see, e.g., \citealp{deutscher2019demand}).

\begin{table}[htb]
\centering
\caption{Parameter estimates with 95\% confidence intervals for the model including covariates in the state-dependent process. 
}\vspace{0.3em}
\label{tab:resultscovariates1}
\begin{tabular}{lccc}
  \hline
 parameter & estimate & \multicolumn{1}{c}{95\% CI} \\ 
  \hline
$\phi$ & 0.983 & [0.982;\,0.984] \\ 
$\sigma_g$ & 0.192 & [0.188;\,0.196] \\
$\sigma$ & 0.873 & [0.871;\,0.874] \\
\hdashline
$\alpha_0$ & 0.842 & [0.682;\,1.002] \\
$\alpha_1$ $\big(eloteam\big)$ & 0.439 & [0.393;\,0.485] \\
$\alpha_2$ $\big(eloopp\big)$ & -0.255 & [-0.300;\,-0.211] \\
$\omega_1$ $\big(friday\big)$ & 0.143 & [-0.007;\,0.353] \\
$\omega_2$ $\big(saturdayaft\big)$ & -0.260 & [-0.431;\,-0.088] \\
$\omega_3$ $\big(saturdayeve\big)$ & 0.004 & [-0.182;\,0.191] \\
$\omega_4$ $\big(sunday\big)$ & 0.006 & [-0.207;\,0.218] \\
\hdashline
$\gamma_0$ & -4.438 & [-4.650;\,-4.225] \\
$\gamma_1$ $\big(gini\big)$ & -10.054 & [-10.718;\,-9.391] \\
$\gamma_2$ $\big(gini^2\big)$ & 14.605 & [14.083;\,15.126] \\
$\gamma_3$ $\big(int\big)$ & -0.542 & [-0.779;\,-0.305] \\
$\gamma_4$ $\big(gini \cdot int\big)$ & 3.638 & [2.938;\,4.338] \\
$\gamma_5$ $\big(gini^2 \cdot int\big)$ & -4.291 & [-4.814;\,-3.767] \\
   \hline
\end{tabular}
\end{table} 

As the estimated effects on $\hat{\pi}_t$, i.e.\ the probability to observe no stakes during an open market in interval $t$, include quadratic terms and interaction effects, Figure \ref{img:interactions} in the Supplementary Material provides a three-dimensional plot of the estimated effects to aid interpretation. 
We find that bettors prefer rather uncertain matches, as the estimated probability for stakes equals to 0 increases with an increasing Gini coefficient, especially at the end of a match. 

\begin{table}[htb]
\centering
\caption{Parameter estimates with 95\% confidence intervals for the final model. 
}\vspace{0.3em}
\label{tab:resultscovariates_state}
\begin{tabular}{lccc}
  \hline
 parameter & estimate & \multicolumn{1}{c}{95\% CI} \\ 
  \hline
$\phi$ & 0.968 & [0.967;\,0.970] \\ 
$\sigma_g$ & 0.202 & [0.198;\,0.207] \\
$\sigma$ & 0.870 & [0.869;\,0.872] \\
\hdashline
$\alpha_0$ & 1.097 & [0.998;\,1.197] \\
$\alpha_1$ $\big(eloteam\big)$ & 0.382 & [0.353;\,0.411] \\
$\alpha_2$ $\big(eloopp\big)$ & -0.330 & [-0.358;\,-0.302] \\
$\omega_1$ $\big(friday\big)$ & 0.007 & [-0.060;\,0.205] \\
$\omega_2$ $\big(saturdayaft\big)$ & -0.209 & [-0.314;\,-0.105] \\
$\omega_3$ $\big(saturdayeve\big)$ & 0.287 & [0.155;\,0.419] \\
$\omega_4$ $\big(sunday\big)$ & 0.006 & [-0.053;\,0.176] \\
\hdashline
$\beta_1$ $\big(halftime\big)$ & 0.002 & [0.000;\,0.005] \\
$\beta_2$ $\big(\textit{ginidiff}\big)$ & -0.094 & [-0.099;\,-0.089] \\
$\beta_3$ $\big(\textit{goalteamsurprising}\big)$ & 0.531 & [0.494;\,0.567] \\
$\beta_4$ $\big(\textit{goalteamslightlysurprising}\big)$ & 0.218 & [0.176;\,0.261] \\
$\beta_5$ $\big(\textit{goalteamunsurprising}\big)$ & 0.034 & [0.001;\,0.067] \\
   \hline
\end{tabular}
\end{table}

\subsection{Covariates in the state process}
For the final model formulation, i.e.\ the model including news covariates in the state process, the results are displayed in Table \ref{tab:resultscovariates_state}. For the covariates in the state process, we observe that the estimated coefficient $\hat{\beta}_1$ suggests that the market activity level is slightly increased during halftime. This result is in line with the findings by \citet{croxson2014information}, who also found an elevated level of betting activity during halftime break. 
Regarding the effect of news, the coefficient $\hat{\beta}_2$ --- which measures the effect of the change in the Gini coefficient compared to the beginning of the match (\textit{ginidiff}) --- indicates that the market activity level increases for unexpected courses of a match. This effect coincides with our prior intuition built based on the example match shown in Figure \ref{img:grafik-dummy}.

We further observe that the effect of a scored goal depends on the odds, and thus on the probability to win the match, right before the goal. A surprising goal scored by the team considered tends to increase the market activity level and hence the amount of stakes placed on the team to have scored ($\hat{\beta}_3 = 0.531$). In contrast, this effect is only half as large for slightly surprising goals ($\hat{\beta}_4 = 0.218$), and vanishes almost completely for unsurprising goals ($\hat{\beta}_5 = 0.034$). While, in general, positive effects of goals may seem intuitive, it is worth remembering that the odds (and hence the potential profits) for the goal-scoring team naturally decrease. Our results thus indicate that the increased confidence of bettors regarding the outcome more than outweighs the substantially decreased odds, especially for surprising goals.

The estimated effects of these different kinds of news on betting behaviour can to some extent be explained by well-known cognitive biases. A well-documented finding in psychology and behavioural economics is that humans tend to ``overreact'' to (unexpected) events and news (see, e.g., \citealp{tang,extrem}) and to attach too much weight to recent information (see, e.g. \citealp{durand}). For stock markets, such overreaction of investors results in prices moving too heavily compared to shifts in fundamental data (see, e.g., \citealp{de1985does}; \citealp{howe1986evidence}; \citealp{kudryavtsev2018availability}). 
In our analysis, we observe an increased market activity level for unexpected courses of a match ($\hat{\beta}_2$) as well as surprising and slightly surprising goals ($\hat\beta_3 > 0$ and $\hat\beta_4 > 0$ ), which might be due to overreaction of bettors. 
Bettors appear to overreact in particular to goals, as the market activity level for the goal-scoring team increases even though the associated decreased odds will obviously reflect these news. 
Bookmakers could exploit this bias by increasing their margin for the odds of the goal-scoring team, as they can still expect clients to place bets on that team to win the match. 
Following such a pricing strategy could thus increase the bookmakers' profits.

\subsection{Model checking}

\begin{figure}[!htb]
	\centering
	\includegraphics[scale=0.5]{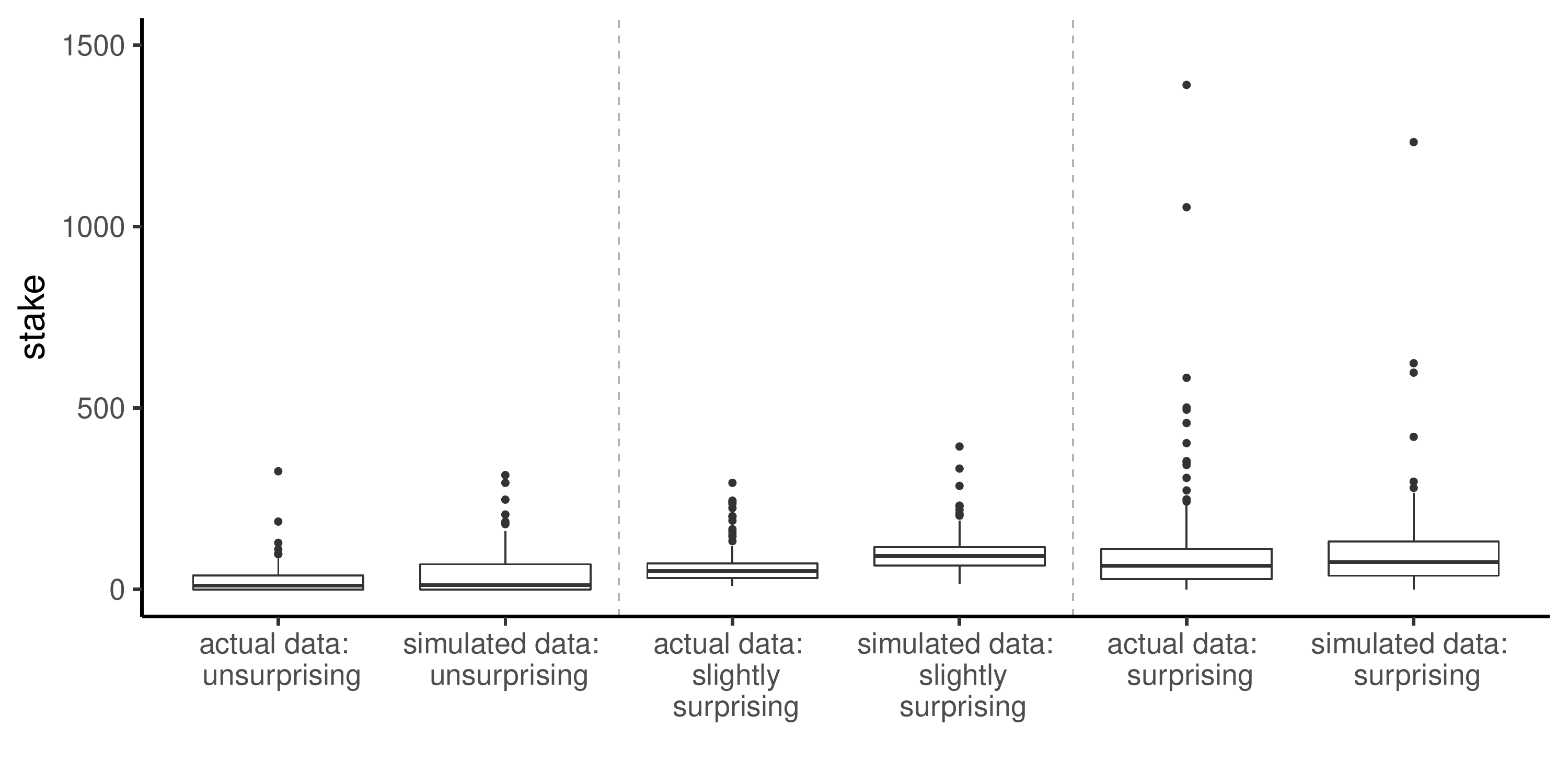}
	\caption{Simulation-based model check by comparing the stakes placed in the actual data to the corresponding stakes placed as obtained in 500 simulation runs, separated for the different kind of news (i.e.\ goal categories).}
	\label{ACF}
\end{figure}

To check the adequacy of our final model, we compare the actual stakes observed in our data to stakes as expected under our model, which are obtained using Monte Carlo simulation. 
Figure \ref{wob} in the Supplementary Material shows the observed stakes for one example match found in the data together with one simulated time series.
To obtain a more detailed picture of our final model aside from such visual inspections, we compare the actual stakes to the corresponding simulated stakes for all goals scored in all matches. In particular, to check whether our model mimics the betting patterns after scored goals, for each goal scored we consider the sum of the actual stakes placed during the first three minutes after a goal. For the simulated data, we average the simulated stakes for each three-minute interval over 500 Monte Carlo simulations. To further check the adequacy of our model with respect to the different types of news, we again distinguish between unsurprising goals, slightly surprising goals, and surprising goals.
The comparison is shown in Figure~\ref{ACF}, demonstrating that our final model adequately captures the stakes placed after a goal. 
The most notable mismatch is found for the slightly surprising goals, where the stakes of the actual data are smaller than expected by our model.

A further characteristic of our data is that sometimes no stakes are placed, e.g.\ in situations where the match is almost decided. To check whether our model is able to capture the occurrence of these zeros, we compare the proportion of observations where no stakes are placed to the expected proportion under our final model. For the model-based proportion, we again use Monte Carlo simulation. Comparing these proportions over all matches, we find that our model predicts that no stakes were placed in about 7.3\% of all intervals, which is very close to the true proportion of 7.2\%. In addition, our model correctly predicts 96.5\% of the intervals where stakes were placed.

\section{Discussion}

In this contribution, we considered a unique data set on stakes placed in a football live betting market to investigate the market's response to news. Using state-space models, we find that bettors are attracted by outcome uncertainty, as our results indicate that the underlying market activity level is increased for such situations. For the main type of news in a football match, i.e.\ goals, we also find an increased market activity level. However, the magnitude of this increase depends on how much of a surprise is represented by the news. While the effect of unsurprising goals on the market activity is relatively small, we observe a notably elevated market activity level following surprising goals. While perhaps intuitive, this finding could not necessarily be expected, as the betting odds for the goal-scoring team should already incorporate the new information, thus counter-balancing the increased confidence in the outcome. The finding may potentially be caused by overconfidence, or overreaction, of bettors, driven by the positive news. Such overconfidence has been intensively studied in psychology and economics, where it has been shown that overconfidence leads to overreaction of investors to recent news. 

The current analysis does not allow for inference on how \textit{individual} bettors react to news. With corresponding data on individual bets placed, one could investigate heterogeneity across bettors in the response to news. 
A further limitation of the current approach concerns the modelling approach that was implemented.
Throughout this contribution, the underlying state process --- serving as a proxy for the underlying market activity level --- may affect not only the stakes placed on the team analysed but also the stakes placed on the opposing team, and vice versa. Corresponding dependencies could be modelled as a bivariate time series for each match, rather than two distinct univariate time series. When formulating the state process, the AR(1) process would then be replaced by a vector autoregressive process.

Building on the models developed here, future research could focus on more specific investigations into the live betting market, for example with respect to fraud detection. In recent years, there were several match-fixing scandals, rendering match fixing a growing threat to the integrity of sports. Using the models developed in this contribution, outlier detection based on residuals could be used to uncover potential match fixing. For pre-game betting, such an approach was developed for example in \citet{matchfixing} and \citet{forrest2019using}. 
For such an outlier detection in the live betting market, it would seem to be crucially important to incorporate the underlying market activity level, as we propose with the SSM approach considered here. For example, if news such as goals lead to a notably increased market activity level in a certain period of time, associated with very large stakes, then a model without a corresponding state process would --- most likely erroneously --- label these stakes as suspicious. 
To avoid such false positives in an automated system for detecting match fixing, it is thus important to explicitly model the general market activity level and how it evolves over time driven by news during the match.

\FloatBarrier
\newpage

\section*{Supplementary Material}
\begin{figure}[htb]
	\centering
	\includegraphics[scale=0.6]{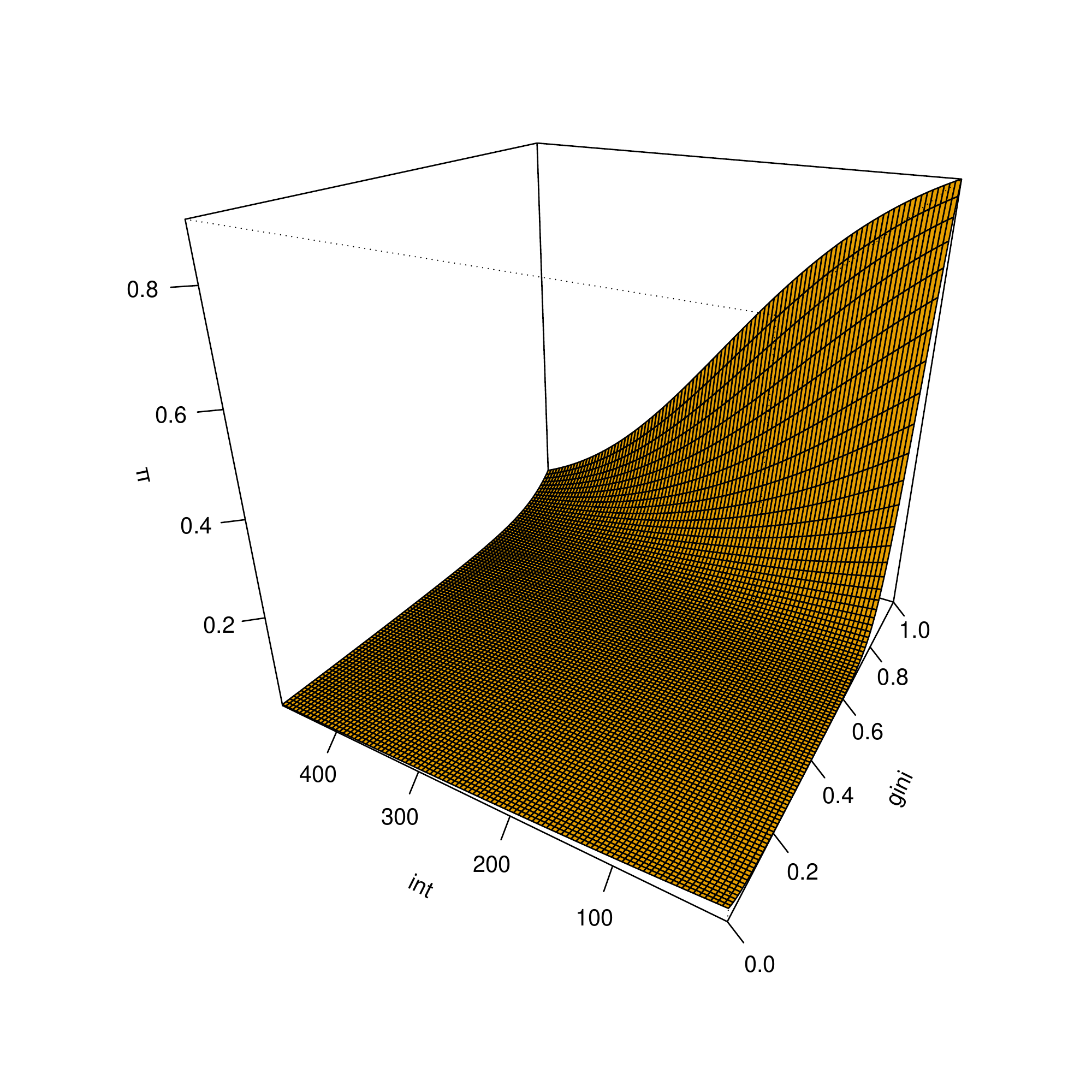}
	\vspace{-1cm}
	\caption{Estimated interaction effects of \textit{int} and \textit{gini} for the model with covariates in the state-dependent process.}
	\label{img:interactions}
\end{figure}

\begin{figure}[htb]
	\centering
	\includegraphics[scale=0.5]{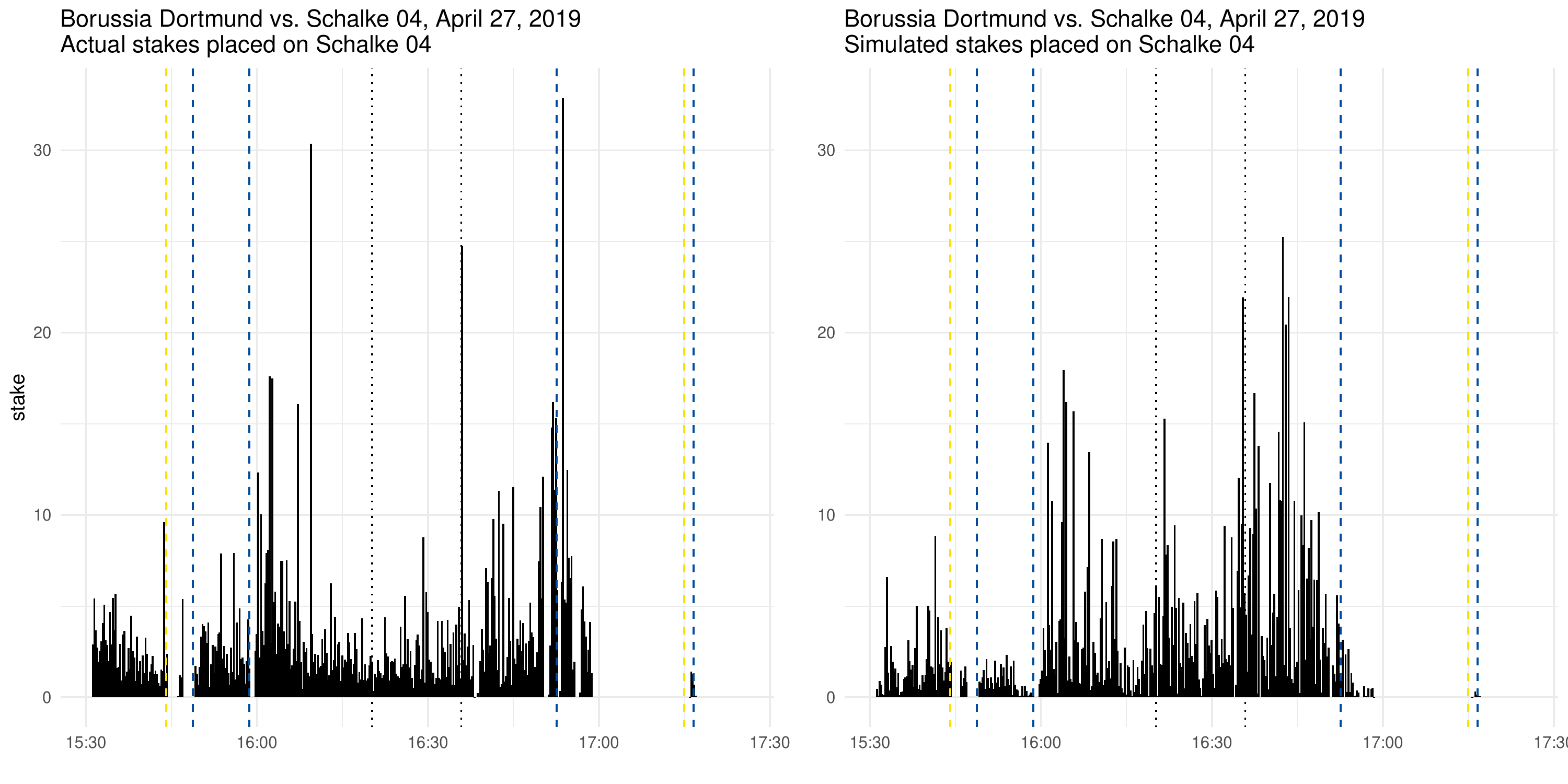}
	\caption{Example time series of stakes found in our data from the match Borussia Dortmund vs.\ Schalke 04. Actual stakes (placed on Schalke 04) are shown on the left, whereas the right panel shows a corresponding simulated time series. The yellow dashed lines denote goals scored by Borussia Dortmund and the blue dashed lines indicate goals scored by Schalke 04.}
	\label{wob}
\end{figure}

\end{spacing}
\end{document}